%Paper: 9108026
%From: <SNOJIRI%FNAL.BITNET@uicvm.uic.edu>
%Date: Thu, 29 Aug 1991 11:38 CDT
%Date (revised): Sun, 1 Sep 1991 22:50 CDT

\input phyzzx

\titlepage
\hskip 7cm FERMILAB-PUB-91/230-T

\hskip 7cm KEK preprint 91-99, KEK-TH 388

\hskip 7cm OCHA-PP-18

\vfil

\centerline{\bf Superstring in Two Dimensional Black Hole}

\vfil

\author{Shin'ichi Nojiri\foot{On leave of absence from National Laboratory for
High Energy Physics (KEK), Oho 1-1, Tsukuba-shi, Ibaraki-ken 305, JAPAN
and Ochanomizu University, 1-1 Otsuka 2, Bunkyo-ku, Tokyo 112, JAPAN.
e-mail address : NOJIRI@JPNKEKVX, NOJIRI@JPNKEKVM}}

\vfil

\centerline{\it Theory Group, Fermi National Accelerator Laboratory}
\centerline{\it P.O.Box 500, Batavia, IL60510, USA}
\vfil
\centerline{\bf abstract}

We construct superstring theory in two dimensional black hole
background based on supersymmetric $SU(1,1)/U(1)$ gauged Wess-Zumino-Witten
model.

\endpage

\REF\ri{E. Witten\journal Phys.Rev.&D44 (91) 314}
\REF\rii{R. Dijkgraaf, H. Verlinde and E. Verlinde, preprint PUPT-1252,
IAASNS-HEP-91/22}
\REF\rxii{S.B. Giddings and A. Strominger, preprint UCSBTH-91-35}
\REF\rxi{G.T. Horowitz and A. Strominger\journal Nucl.Phys. &B360 (91) 197}
\REF\rv{M.M. Nojiri and S. Nojiri\journal Prog.Theor.Phys. &B83 (90)677}
\REF\riv{Y. Kazama and H. Suzuki\journal Nucl.Phys. &B321 (89)232}
\REF\riii{L.J. Dixon, J. Lykken and M.E. Peskin\journal Nucl.Phys.&B325 (89)
326}
\REF\rvi{E. Marinari and G. Parisi\journal Phys.Lett. &B240 (90)561}
\REF\rvii{S. Nojiri\journal Phys.Lett. &B252(90)561
 \journal Phys.Lett. &B253 (91)
63 \journal Prog. Theor.Phys. &85 (91) 671}
\REF\rviii{S. Nojiri\journal Phys.Lett. &B262 (91)419 \journal Phys.Lett.
&B264 (91)57}
\REF\rix{P. Di Vecchia, V.G. Knizhnik, J.L. Petersen and P. Rossi\journal
Nucl.Phys. &B253 (85)701}
\REF\rxviii{D.Z. Freedman and P.K. Townsend\journal Nucl.Phys.
&B177 (81) 282}
\REF\rxv{T. Kugo and I. Ojima\journal Phys.Lett. &B73 (78) 459
\journal Prog.Theor.Phys. Supp. &66 (79) 1}
\REF\rxvi{M. Ito, T. Morozumi, S. Nojiri and S. Uehara\journal
Prog.Theor.Phys. &75 (86) 934}
\REF\rxvii{N. Ohta\journal Phys.Rev. &D33 (86) 1681 \journal Phys.Lett.
&B179 (86) 347}
\REF\rxiii{N. Ishibashi, M. Li and A.R. Steif, preprint UCSBTH-91-28 (Revised
Version)}
\REF\rxiv{J.H. Horne and G.T. Horowitz, preprint UCSBTH-91-39}
\REF\rx{P. Di Vecchia, J.L. Petersen and H.B. Zheng\journal Phys.Lett.
&B174 (86)280}

Recently it was shown that the $SU(1,1)/U(1)$ gauged Wess-Zumino-Witten
(GWZW) model describes strings in a two dimensional black
hole.\refmark\ri
The string propagation and Hawking radiation in this black hole were
discussed in Ref.\rii .
When the level $k$ of $SU(1,1)$ current algebra equals to $9/4$, this model
can be regarded as a two dimensional gravity coupled with $c=1$
superconformal matter.
We expect that
this model could be one of toy models which provide a clue to solve
the dynamics of more \lq\lq realistic" string models.

The supersymmetric extension of this model appeared\refmark\rxii
as an exact solution of
ten-dimensional superstring theory corresponding to black
fivebranes.\refmark\rxi
In this paper, we will consider the supersymmetric extension
based on $SU(1,1)/U(1)$ supersymmetric GWZW (SGWZW) model.
It has been shown\refmark\rv that supersymmetric $SU(1,1)/U(1)$ coset model
has $N=2$
supersymmetry due to Kazama-Suzuki\refmark\riv mechanizm and this model is
equivalent to $N=2$ superconformal models proposed by Dixon, Lykken and
Peskin.\refmark\riii
The central charge $c$ of this system is given by,
$$c={3k \over k-2} \ . \eqn\abi$$
Here $k$ is the level of $SU(1,1)$ current algebra. When $k={5 \over 2}$,
the central charge $c$ equals to 15 and this conformal field theory
describes a critical string theory.
The $N=1$ supergravity coupled with $c={3 \over 2}\hat c={3 \over2}$
superconformal matter would be described by this critical theory.
Furthermore the $N=2$ superconformal symmetry of this model suggests
that pure $N=2$ supergravity would be also described by this model
when $c=6$ ($k=4$).
Due to $N=2$ superconformal symmetry, superstring theories in the two
dimensional black hole background can be constructed by imposing GSO
projection.
It is also expected that this superstring theory
would be equivalent to the matrix models which have space-time
supersymmetry\refmark{\rvi, \rvii}
and topological superstring theories based on $N=2$ superconformal topological
field theories.\refmark\rviii

The action $S^{\rm WZW}(G)$ of $N=1$ supersymmetric Wess-Zumino-Witten model
is given by,\refmark\rix
$$\eqalign{S^{\rm WZW}(G)
=&{k \over 2\pi}\int d^2z d^2\theta {\rm tr}G^{-1}DGG^{-1}\bar DG
\cr
&-{k \over 2\pi}\int dt d^2z d^2\theta
[{\rm tr}G^{-1}DGG^{-1}\bar DG G^{-1}\partial_t G
 +(D \leftrightarrow \bar D {\rm term})] \ . }\eqn\i$$
Here we define
covariant derivatives $D$ and $\bar D$
by using holomorphic and anti-holomorphic Grassmann
coordinates $\theta$  and $\bar \theta$
$$D\equiv {\partial \over \partial \theta}-
\theta{\partial \over \partial z} \ , \ \ \ \
\bar D\equiv {\partial \over \partial \bar \theta}-
\bar \theta{\partial \over \partial \bar z}\ . \eqn\ii$$
The matrix superfield $G$, which is an element of a group {\cal G},
is given by
$$G=\exp (i\sum_a T^a \Phi^a)\ .\eqn\iii$$
Here $\Phi^a$ is a superfield $\Phi^a=\phi^a+\theta\psi^a
+\bar\theta\,\bar\psi^a+\theta\bar\theta f^a$ and $T^a$ is a generator of
the algebra corresponding to {\cal G}.
The action \i\ satisfies Polyakov-Wiegmann type formula:
$$S^{\rm WZW}(GH)=S^{\rm WZW}(G)+S^{\rm WZW}(H)+{k \over \pi}
\int d^2z d^2\theta {\rm tr}G^{-1}DG \bar DH H^{-1}\ .\eqn\iv$$
Here $H$ is also an element of {\cal G}.
This formula guarantees that the system described by the action \i\ has
super Kac-Moody symmetry and $N=1$ superconformal symmetry.

If {\cal F} is a $U(1)$ subgroup of {\cal G}, we can gauge the global symmetry
under the following axial $U(1)$
transformation, which is given by an element $F$ of {\cal F},
$$G \rightarrow FGF\ . \eqn\biii$$
The action of supersymmetric ${\cal G}/{\cal F}$ gauged Wess-Zumino-Witten
is given by,
$$\eqalign{S^{{\cal G}/{\cal F}}(G, A)=&S^{\rm WZW}(F_LGF_R)-
S^{\rm WZW}(F_L^{-1}F^R)\cr
=&S^{\rm WZW}(G) \cr
&+{k \over 2\pi}\int d^2z d^2\theta {\rm tr}(A\bar A+AG\bar A G^{-1}+
G^{-1}DG\bar A+A \bar DG G^{-1}) \ .} \eqn\v$$
Here $F_L$ and $F_R$  are elements of {\cal F} and gauge fields $A$ and
$\bar A$ are defined by,
$$A=F_L^{-1}DF_L , \ \ \ \ \bar A=F_R^{-1}\bar DF_R\ . \eqn\vi$$
The action \v\ is invariant under the following $U(1)$ gauge transformation
$$\eqalign{G \rightarrow FGF , \ \ A \rightarrow A&+F^{-1}DF ,
\ \ \bar A \rightarrow \bar A+ F^{-1}\bar DF \ . \cr
&(F_{L,R} \rightarrow F_{L,R}F) }\eqn\biv$$

A supersymmetric extension of string theory in a two dimensional black hole
background is given by setting ${\cal G}=SU(1,1)$ in the action \v .
We start with considering $SU(1,1)$ SWZW model.
By parametrizing $G$ by,
$$G=\exp ({i \over 2}\Phi_L\sigma_2) \exp ({1 \over 2}R\sigma_1)
\exp ({i \over 2}\Phi_R\sigma_2) , \eqn\vii$$
with $\sigma_i$ the Pauli matrices,
we obtain the action $S^{SU(1,1)}$ of $SU(1,1)$ SWZW model
$$\eqalign{S^{SU(1,1)}=&{k \over 2\pi}\int d^2z d^2\theta
[-{1 \over 2}D\Phi_L\bar D\Phi_L
-{1 \over 2}D\Phi_R\bar D\Phi_R \cr
&-{\rm cosh}R\,D\Phi_L\bar D\Phi_R+{1 \over 2}DR\bar DR]\ .}\eqn\ai$$
The holomorphic (anti-holomorphic) conserved currents $J_i$ ($\bar J_i$)
of this system are given by,
$$2kG^{-1}DG=J_1 \sigma_1 +iJ_2 \sigma_2 +J_3 \sigma_3\ , \ \
2kG^{-1}\bar DG=\bar J_1 \sigma_1 +i\bar J_2 \sigma_2
+\bar J_3 \sigma_3\ , \eqn\xxv$$
$$J_i=j_i + \theta{\tilde J_i}+\cdots \ ,\ \ \
\bar J_i=\bar j_i + \bar \theta \bar{\tilde J_i}+ \cdots \ . \eqn\xxvi$$
Here $\cdots$ express the terms which vanish by using
the equations of motion.
If we define new currents $\hat J_i$ and $\bar{\hat J}_i$ by the following
equation
$$\hat J_i=\tilde J_i-{1 \over 2k}\epsilon_{ilm}j_lj_m , \ \
\bar{\hat J}_i=\bar{\tilde J}_i
-{1 \over 2k}\epsilon_{ilm}\bar j_l\bar j_m , \eqn\aii$$
These currents $\hat J_i$ and $\bar{\hat J}_i$ do not depend on the fermion
currents $j_i$ and $\bar j_i$.

By expanding superfields $\Phi_{L,R}$ and $R$ into components,
$$\eqalign{
\Phi_{L,R}&=\phi_{L,R}+\theta\psi_{L,R}+\bar\theta\,\bar\psi_{L,R}+
\theta\bar\theta f_{L,R} , \cr
{R \over 2}&=s+\theta\eta+\bar\theta\bar\eta+\theta\bar\theta g ,}
\eqn\bx$$
we can rewrite the $SU(1,1)$ SWZW action $S^{SU(1,1)}$ in
Eq.\ai\ by a sum of non-supersymmetric
$SU(1,1)$ WZW action $\tilde S^{SU(1,1)}$
and free fermion actions:
$$\eqalign{S^{SU(1,1)} =&\tilde S^{SU(1,1)}
+{1 \over 4k\pi}\int d^2z [j_+\bar\partial j_--j_2\bar\partial j_2
+\bar j_+\partial \bar j_--\bar j_2\partial \bar j_2]\ , }\eqn\aiii$$
$$\eqalign{\tilde S^{SU(1,1)}=&{k \over 2\pi}
\int d^2z [-{1 \over 2}(\partial\phi_L\bar\partial\phi_L
+\partial\phi_R\bar\partial\phi_R)\cr
&-{\rm cosh}(2s) \partial\phi_L\bar\partial\phi_R +2\partial s\bar\partial s]
\ .}
\eqn\xxviii$$
Here $j_\pm$ and $\bar j_\pm$ are defined by
$$j_\pm\equiv j_1\pm ij_3\ , \ \ \ \bar j_\pm\equiv \bar j_1\pm i\bar j_3
\ .\eqn\xxix$$
The conserved currents corresponding to the non-supersymmetric
$SU(1,1)$ WZW action $\tilde S^{SU(1,1)}$ \xxviii\ are given by
$\hat J_i$ and $\bar{\hat J}_i$ in Eq.\aii .

Fermionic currents $j_\pm$ and $\bar j_\pm$ can be written as
$$\eqalign{j_\pm =&{k \over2}
\exp (\mp i \phi_R) (\eta\pm{i \over 2}{\rm sinh}(2s)\psi_L)\  ,
\cr
\bar j_\pm =&{k \over 2}\exp (\mp i \phi_L)
(\bar\eta\pm{i \over 2}{\rm sinh}(2s)\bar\psi_R) \ .} \eqn\bi$$
Note that there appear bosonic factors
$\exp (\mp i \phi_R) $ and $\exp (\mp i \phi_L) $.
Due to these factors, the boundary conditions of
$j_\pm$ and $\bar j_\pm$ are twisted although fermions $\eta$, $\bar\eta$,
$\psi_L$ and $\bar\psi_R$, which is identified later with
space-time fermionic coordinates,
are periodic or anti-periodic. Therefore the eigenvalues of the zero modes
of fermion number currents $K$ and $\bar K$,
$$K={1 \over 4k}(j_+j_--j_-j_+)\ , \ \ \
\bar K={1 \over 4k}(\bar j_+\bar j_--\bar j_-\bar j_+)\ , \eqn\xxx$$
which satisfy the following operator product expansions
$$K(z)j_\pm(w)\sim\pm{1 \over z-w}j_\pm\ , \ \ \
\bar K(\bar z)\bar j_\pm(\bar w)\sim\pm{1 \over \bar z-\bar w}
\bar j_\pm\ , \eqn\xxxi$$
are not quantized.

We now gauge the $U(1)$ symmetry in the action \ai\ by following Eq. \v .
We consider the case that the $U(1)$ symmetry is generated by
$\sigma_2$. Since the $U(1)$ symmetry is
compact, the resulting conformal field theory
describes the Euclidean black hole.
The theory of the Lorentzian black hole can be obtained by replacing
$\sigma_2$ by $\sigma_3$ or simply by analytic continuating
$\Phi_{L,R}\rightarrow i\Phi_{L,R}$.

By the parametrization \vii\
the $SU(1,1)/U(1)$ gauged SWZW action
takes the form
$$\eqalign{S^{SU(1,1)/U(1)}=&S^{SU(1,1)}
+{k \over 2\pi}\int d^2z d^2\theta [4(1+{\rm cosh}R)A\bar A \cr
&+2iA(\bar D \Phi_L+{\rm cosh}R\,\bar D \Phi_R)
+2i(D \Phi_L+{\rm cosh}R\, D \Phi_R)\bar A] \ .}\eqn\viii$$
Here $S^{SU(1,1)}$ is $SU(1,1)$ SWZW action in Eq.\ai .
By using the following redefinitions,
$$\eqalign{\Phi\equiv &\Phi_L-\Phi_R\ , \cr
A'\equiv &A+{i \over 2}{{\rm cosh}R\,D\Phi_L+D\Phi_R \over 1
+ {\rm cosh}R}\ , \cr
\bar A'\equiv &\bar A
+{i \over 2}{{\rm cosh}R\,\bar D\Phi_R+\bar D\Phi_L \over 1 + {\rm cosh}R}\ , }
\eqn\ix$$
the action \viii\ can be rewritten as follows,
$$\eqalign{S^{SU(1,1)/U(1)}={k \over 2\pi}&\int d^2z d^2\theta
[{1 \over 2}{\rm tanh}^2{R \over 2}\,D\Phi\bar D\Phi \cr
&+{1 \over 2}DR\bar DR+4(1+{\rm cosh}R)A'\bar A' \ . }\eqn\x$$
The action \viii\ and \x\ are invariant under the following
infinitesimal gauge transformation corresponding to Eq.\biv ,
$$\delta \Phi_L=\delta \Phi_R=\Lambda \ , \ \
\delta A=-{i \over 2}D\Lambda \ , \ \
\delta \bar A=-{i \over 2}\bar D\Lambda \ . \eqn\xi$$
We fix this gauge symmetry by imposing the gauge condition
$$\Phi_L=-\Phi_R=\tilde \Phi \ . \eqn\xii$$
By integrating gauge fields $A$ and $\bar A$ in
the action \viii\ or \x ,\foot{
The integration of the gauge fields induces the dilaton term in the action
but we now neglect this term. The gauge fixed action which is correct
at the quantum level is given later in this paper.}
and by integrating auxiliary fields,
we obtain the following action,
$$\eqalign{
S^{(1)}={k \over \pi}&\int d^2 z
[{\rm tanh}^2 s (\partial \phi \bar \partial \phi
- \partial \bar \psi \,\bar \psi
+\psi \bar \partial \psi) \cr
&-2{{\rm sinh} s \over {\rm cosh}^3 s}(\eta\psi\bar\partial \phi
+\bar\eta\bar\psi\partial \phi)
+4\,{\rm tanh}^2 s\, \eta\bar\eta\psi\bar\psi \cr
&+\partial s \bar \partial s - \partial \bar \eta \,\bar \eta
+\eta \bar \partial \eta] \ .}\eqn\xiii$$
Here we write superfields $\tilde\Phi$ and $R$ in terms of components:
$$\eqalign{
\tilde\Phi&=\phi+\theta\psi+\bar\theta\,\bar\psi+\theta\bar\theta f
\ , \cr
{R \over 2}&=s+\theta\eta+\bar\theta\bar\eta+\theta\bar\theta g}\ .
\eqn\xii$$
This system has $N=1$ supersymmetry since the starting action \viii\
and gauge condition \xi\ are manifestly supersymmetric.
In fact, this action is nothing but the action of
(1,1) supersymmetric $\sigma$ model\refmark\rxv in two dimensional
black hole background.

The $N=1$ supersymmetry in the action \xiii\
is extended to $N=2$ supersymmetry since this action is invariant
under the following holomorphic
(anti-holomorphic) $U(1)$ symmetry:
$$\eqalign{\delta\psi =&-{u(z) \over {\rm tanh}s} \eta \cr
\delta\eta =&u(z) {\rm tanh}s \,\psi \ ,}\eqn\xv$$
$$\eqalign{\delta\bar\psi =&-{\bar u(\bar z) \over {\rm tanh}s} \bar
\eta \cr
\delta\bar\eta =&\bar u(\bar z) {\rm tanh}s \,\bar\psi\ .}\eqn\xvi$$
Here $u(z)$ ($\bar u(\bar z)$) is a holomorphic (anti-holomorphic)
parameters of the transformation.
The transformations \xv\ and \xvi\ tell that the currents of
this $U(1)$ symmetry
can be regarded as fermion number currents
with respect to
space-time fermion coordinates, $\eta$, $\psi$, $\bar\eta$ and $\bar\psi$.
By commuting this $U(1)$ symmetry transformation with the original $N=1$
supersymmetry transformation, we obtain another supersymmetry transformation
and we find that the action has $N=2$ supersymmetry. On the other hand,
in case of the Lorentzian black hole, the obtained algebra is not exactly
$N=2$ superconformal algebra.\foot{
Note that any Lorentzian manifold is not K\"ahler.}
Usual $N=2$ superconformal algebra is given by
$$\eqalign{\{G_n^+, G_m^-\}=&4L_{n+m}+2(m-n)J_{n+m}+{c \over 12}m(m^2-1)
\delta_{m+n,0} \ , \cr
[J_n, G_m^\pm]=&\pm G_{n+m}\ , \ \ \ \ {\rm etc.}}
\eqn\xvii$$
and the hermiticities of the operators are assigned by
$$(G_n^+)^\dagger =G_{-n}^-\ , \ \ \ J_n^\dagger =J_{-n}\ . \eqn\xviii$$
The algebra which appears in the Lorentzian case is identical with Eq.\xvii ,
but the assignment of the hermiticities is different from Eq.\xviii :
$$(G_n^+)^\dagger =G_{-n}^+\ , \ \ \ (G_n^-)^\dagger =G_{-n}^-\ , \ \ \
J_n^\dagger =-J_{-n} \ . \eqn\xix$$
This is not so surprizing since this algebra also appears in flat two
dimensional Lorentzian space-time which is a subspace of
flat ten dimensional space-time in usual Neveu-Schwarz-Ramond model.
Even in Lorentzian case, we have a $U(1)$ current
and superstring theories can be constructed by imposing
GSO projection.

In order to consider the spectrum of this theory, we choose the following
gauge condition instead of Eq. \xi\ ,
$$\bar DA-D\bar A=0 \ . \eqn\xx$$
This gauge conditon allows us to parametrize the gauge fields $A$ and $\bar A$
as
$$A=D\Pi \ , \ \ \ \bar A=-\bar D\Pi \ . \eqn\xxi$$
By shifting the fields $\Phi_{L,R}$,
$$\Phi_L \rightarrow \Phi_L +2i\Pi \ ,
\ \ \Phi_R \rightarrow \Phi_R -2i\Pi \ ,
\eqn\xxii$$
the gauge fixed action $S^{(2)}$
is given by a sum of $SU(1,1)$ SWZW action $S^{SU(1,1)}$ in
Eq. \ai , free field action $S^{\Pi}$
and (free) ghost action $S^{\rm FP}$.
$$\eqalign{S^{(2)}=&S^{SU(1,1)}+S^{\Pi}+S^{\rm FP}\ , \cr
S^{\Pi}=&-{4k \over \pi}\int d^2z d^2\theta D\Pi \bar D\Pi\ , \cr
S^{\rm FP}=&{k \over 2\pi}\int d^2z d^2\theta BD \bar DC\ . }\eqn\xxiii$$
Here $B$ and $C$ are anti-ghost and ghost superfields.

The BRS charge
$Q_{\rm B}$
which defines the physical states is given by
$$Q_{\rm B}=\oint dz C(D\Pi-{i \over 4k}J_2)+\oint d\bar z
C(\bar D\Pi+{i \over 4k}\bar J_2) \ . \eqn\xxiv$$
This BRS charge gives constraints on the physical states,
$$D\Pi-{i \over 2}J_2=\bar D\Pi+{i \over 2}\bar J_2=0 \ , \eqn\xxvii$$
which tell that $B$, $C$, $\Pi$ and $J_2$ (or $\bar J_2$) make
so-called \lq\lq quartet" structure\refmark\rxv
similar to the structure which appeared
in the quantization of
Neveu-Schwarz-Ramond model based on BRS symmetry.\refmark{\rxvi , \rxvii}

The action which describes superstring theory in the two dimensional
black hole is simply given by a sum of $SU(1,1)$ WZW action \xxviii ,
free fermion actions \aiii\ and the actions of free superfield
and free ghost and anti-ghost superfields \xxiii . Furthermore
the constraints \xxvii\ imposed by th BRS charge \xxiv\ can be easily
solved with respect to free superfields $\Pi$. Therefore if we can find
the spectrum of the bosonic string in the two dimensional
black hole,\refmark{\ri , \rii} we can also find the spectrum
of this string theory.

The $U(1)$ current, which corresponds to the transformations \xv\
and \xvi\ are given by,\refmark\rv
$$J={-2i \over k-2}\hat J_2+{k \over k-2}K\ , \ \ \ \bar J
={-2i \over k-2}\bar{\hat J}_2+{k \over k-2}\bar K\ . \eqn\xxxii$$
Here $\hat J_2$ and $\bar{\hat J}_2$ are defined by Eq.\aii\ and
fermion number currents $K$ and $\bar K$ are defined by
Eq.\xxx .
These $U(1)$ currents commute with the BRS charge \xxiv\ and we can impose
GSO projection consistently.
Note that GSO projection does not give any constraint on the
representations of $SU(1,1)$ current algebra since
the eigenvalues of the zero modes in the currents $K$ and $\bar K$ are
not quantized although those in $J$ and $\bar J$ are quantized.

Recently the string model based on ${SU(1,1)\times U(1) \over U(1)}$ coset
model was discussed.\refmark{\rxiii,\rxiv} This model describes the strings
in two\refmark\rxiii
or three dimensional\refmark\rxiv
charged black holes. By adjusting the radius of the $U(1)$ boson,
we will obtain $N=2$ superconformal theory with $c>3$\refmark\riii
in the same way as $N=2$ minimal model was constructed from
${SU(2)\times U(1) \over U(1)}$\refmark\rx .  The obtained model should be
equivalent to the model discussed here.

I would like to acknowledge discussions with N. Ishibashi, M. Li,
J. Lykken and A. Strominger. I am also indebted to M. Kato, E. Kiritsis,
A. Sugamoto, T. Uchino and S.-K. Yang for the discussion at the early stage.
I wish to thank the theory groups of SLAC, UC Santa Barbara and Fermilab, where
this work was done, for their hospitalities.
I am grateful to Soryuushi Shougakkai for the financial support.

\endpage
\refout
\bye